\documentclass[letterpaper]{article}
\usepackage[hyphens]{url}
\usepackage{graphicx}
\usepackage{caption}
\usepackage{amsmath}
\usepackage{array}
\usepackage{booktabs}
\usepackage{multirow}
\usepackage{multicol}
\usepackage{tabularx}
\usepackage{epstopdf}
\usepackage{tcolorbox}
\tcbuselibrary{breakable}
\usepackage{amssymb}
\usepackage{listings}
\usepackage{xcolor}
\usepackage{enumitem}
\usepackage{algorithm}
\usepackage{algorithmic}
\usepackage{hyperref}
\usepackage[authoryear]{natbib}
\usepackage[top=1in, bottom=1in, left=1in, right=1in]{geometry}

\lstdefinelanguage{json}{
    basicstyle=\normalfont\ttfamily,
    commentstyle=\color{gray},
    numbers=left,
    numberstyle=\scriptsize,
    stepnumber=1,
    numbersep=8pt,
    showstringspaces=false,
    breaklines=true,
    frame=lines,
    stringstyle=\color{red},
    morestring=[b]",
    morecomment=[l]{//},
    morecomment=[s]{/*}{*/},
}

\lstdefinelanguage{SQL}{
    morekeywords={CREATE,TABLE,PRIMARY,KEY,INT,VARCHAR,DATE,DECIMAL},
    sensitive=true,
    morecomment=[l]{--},
    morecomment=[s]{/*}{*/},
    morestring=[b]',
}
\newcommand*\samethanks[1][\value{footnote}]{\footnotemark[#1]}

\title{\textbf{SCoGen: Scenario-Centric Graph-Based Synthesis of Real-World Code Problems}}
\author{
    \textbf{Xifeng Yao\thanks{~~Co-first authors}}, \textbf{Dongyu Lang\samethanks}, \textbf{Wu Zhang\samethanks}, \textbf{Xintong Guo\samethanks}, \textbf{Huarui Xie}, \\
    \textbf{Yinhao Ni}, \textbf{Ping Liu}, \textbf{Guang Shen}, \textbf{Yi Bai}, \textbf{Dandan Tu\thanks{~~Corresponding author}}, \textbf{Changzheng Zhang\samethanks}\\
    Huawei Technologies Co., Ltd.\\
    \{yaoxifeng, zhang.wu1, terry.guoxintong, tudandan, zhangzhangzheng\}@huawei.com\\
    langdongyu@h-partners.com
}
\date{}

\begin{document}
\maketitle

\begin{abstract}
Significant advancements have been made in the capabilities of code large language models, leading to their rapid adoption and application across a wide range of domains. However, their further advancements are often constrained by the scarcity of real-world coding problems. To bridge this gap, we propose a novel framework for synthesizing code problems that emulate authentic real-world scenarios. This framework systematically integrates domain knowledge, domain skills, and coding skills, all of which are meticulously extracted from real-world programming-related datasets, including Stack Overflow and Kaggle. The extracted elements serve as the foundational building blocks for constructing code problems. To align the generated problems with practical applications, application scenarios are also mined from the aforementioned datasets. These scenarios are then utilized to construct a scenario-centric graph that interconnects domain knowledge, domain skills, and coding skills. Based on this structured representation, a sampling strategy on the graph is designed, which effectively controls the generation of a code problem with complexity and diversity, reflects real-world challenges. Experimental results demonstrate that the proposed method consistently achieves superior performance over state-of-the-art open-source large language models of varying sizes and functionalities, including both coders and general-purpose models, across a diverse set of real-world benchmarks.
\end{abstract}


\section{Introduction}

The rapid evolution of large language models (LLMs) is catalyzing a profound transformation within the domain of software engineering \citep{Haque_2025}. Cutting-edge models, such as Claude Opus 4 \citep{anthropic2025claude4}, OpenAI o3 \citep{openai2025o3o4mini}, and Google's Gemini \citep{deepmind2025gemini25pro}, have exhibited remarkable capabilities in a broad spectrum of code-related tasks, encompassing complex, real-world software engineering problems. In parallel, open-source LLMs such as DeepSeek-R1 \citep{deepseek2025r1} and the Qwen series \citep{yang2025qwen3technicalreport} have shown competitive performance on standardized coding benchmarks. Nevertheless, their efficacy in addressing practical, real-world software engineering challenges remains substantially weaker when compared to their proprietary counterparts. A critical contributing factor lies in the fact that these closed-source models are developed under closed development mode, with limited transparency concerning their training data composition and data curation strategies.

To bridge this gap, open-source initiatives have primarily focused on two complementary directions: (1) enhancing the utility of real-world code data through improvements in data quality \citep{seed2025seedcoderletcodemodel}, and (2) adopting synthesis strategies to generate high-quality synthetic code data \citep{qin2025scalinglawssyntheticdata} \citep{wang2023selfinstructaligninglanguagemodels} \citep{ge2025scalingsyntheticdatacreation} \citep{li2024syntheticdataalmostscratch} \citep{xu2024magpiealignmentdatasynthesis}. Real code data is highly valuable due to its extensive availability and rich diversity in terms of programming languages, coding styles, task domains, software engineering practices etc. More importantly, the sheer volume of real code data is conducive to the scaling laws observed in LLMs, enabling them to achieve better performance with increased training data. Nevertheless, real code data often suffers from significant quality issues,  which make the curation process both time-consuming and resource-intensive \citep{huang2025opencoderopencookbooktoptier} \citep{guo2024deepseekcoderlargelanguagemodel} \citep{Shao2024DeepSeekV2AS} \citep{li2023starcodersourceyou} \citep{qwen2025coder}. As a result, synthesis-based strategies have been widely adopted in the post-training phase for their efficiency, cost-effectiveness, and controllability in generating structured and high-quality code samples. However, existing synthetic code generation methods predominantly focus on function-level or algorithmic tasks, which fail to capture the multifaceted nature of real-world software development that typically spans multiple domains and requires the integration of diverse technical skill, as illustrated in Appendix~\ref{appendix: Real-world vs Algorithm-level}.

To systematically address these interrelated challenges, we propose a novel framework for synthesizing real-world code problems, which fundamentally reimagines the methodology for generating training data for code LLMs. Our approach begins with the systematic extraction of application scenarios, domain knowledge, domain skills, and coding skills from extensive, real-world programming-related datasets, including Stack Overflow \citep{stackoverflow} and Kaggle \citep{kaggle}. We posit that these three distinct yet interdependent dimensions — domain knowledge, domain skills, and coding skills — collectively offer a comprehensive and precise characterization of a code task. Leveraging this multi-dimensional information, we construct more realistic and contextually grounded code problems. To further ensure the fidelity of the generated problems to actual software development project practices, we introduce a scenario-centric graph structure that integrates and interconnects the aforementioned dimensions of knowledge and skills. This graph-based representation enables us to model the relationships between various components of a coding task in a structured and interpretable manner. Based on this representation, we design a sampling strategy that takes into account both the complexity defined by the number of knowledge-skill combinations and the diversity controlled through the use of different temperature parameters in the sampling process. Once the knowledge and skill combinations are sampled, we employ prompt engineering with the Qwen3-32B \citep{yang2025qwen3technicalreport} model to generate the final code problems.

We conduct extensive experiments across two benchmark categories, including real-world-level and algorithm-level tasks. The evaluation results demonstrate that, in comparison to existing instruction-tuned models, our method achieves significant improvements on real-world-level benchmarks. For basic algorithm-level problems, although our training data do not include such tasks, our method remains competitive, suggesting that the our method generalizes effectively to fundamental coding scenarios.

In summary, our contributions are: (1) we propose a systematic strategy for extracting fundamental elements that effectively constitute a code problem; (2) we design a scenario-centric knowledge graph that effectively integrates domain knowledge, domain skills, and coding skills, offering a unified framework for modeling their interdependencies; (3) we introduce a novel sampling strategy that allows for precise control over the complexity and diversity of synthetic code problems; (4) we conduct extensive experiments and ablation studies to comprehensively evaluate the effectiveness and robustness of our proposed approach.

\section{Related Work}
\subsection{Code Data Synthesis}
The performance of code LLMs has seen substantial enhancement through the strategic generation of synthetic data. Early pioneering efforts, such as Code Alpaca \citep{codealpaca} and WizardCoder \citep{xu2025wizardlmempoweringlargepretrained}, introduced the use of LLMs to refine and expand simple seed instructions to enhance the instruction-following capabilities. Subsequent research has largely concentrated on enhancing the diversity of tasks or algorithmic-level problem generation \citep{Liu2025rStarCoderSC}. For example, Magpie \citep{xu2024magpiealignmentdatasynthesis} leverages the Llama-3-Instruct model \citep{grattafiori2024llama3herdmodels} to synthesize diverse user queries, yet this method offers relatively limited control over the problem structure and grounding. In contrast, KodCode \citep{xu2025kodcodediversechallengingverifiable} specifically targets the generation of algorithmic and data structure problems, enriching the pool of classical computational challenges. More recently, EpiCoder \citep{wang2025epicoderencompassingdiversitycomplexity} has proposed a novel feature tree-based synthesis framework that systematically captures hierarchical code features. This approach enables fine-grained control over the complexity of generated code, allowing for the generation of problems ranging from individual functions to multi-file software systems through targeted subtree sampling. Concurrently, Genetic-Instruct \citep{majumdar2025geneticinstructscalingsynthetic} has introduced an evolutionary approach using LLM triads (Instructor-Coder-Judge) to massively scale instruction generation from minimal seeds, demonstrating significant performance gains through synthetic data volume. Notably, models such as OpenCoder \citep{huang2025opencoderopencookbooktoptier} and SeedCoder \citep{seed2025seedcoderletcodemodel} predominantly focus on the generation of instruction-response pairs for isolated, small-scale problems, such as the implementation of individual functions or the correction of localized bugs. 

While these approaches have demonstrably enhanced the performance of code LLMs on algorithm-level benchmarks and basic instruction-following tasks, they exhibit critical limitations: (1) they often lack rich contextual grounding within authentic application scenarios; (2) there is limited explicit control over structural complexity; and (3) the problems seldom reflect the multi-faceted nature of real-world software development requiring the orchestration of diverse competencies. In contrast, our work introduces a paradigm shift by synthesizing comprehensive, scenario-grounded problems that closely mirror real-world software development challenges.

\subsection{Knowledge Graph Representation}
Recent research has increasingly explored the utilization of knowledge graphs for synthetic data generation within the domain of natural language processing. For instance, \citep{qin2025scalinglawssyntheticdata} constructs knowledge graphs based on entities and relations extracted from textual corpora to guide and enhance the text generation, thereby demonstrating the potential benefits of incorporating structured knowledge at scale. However, these graph-based methods often exhibit a limited focus, primarily centering on concepts, while largely overlooking the critical skill dimensions of domain skills, coding skills and crucially, the application scenarios that bind them together meaningfully. 

\section{Methodology}

\begin{figure}
    \centering
    \includegraphics[width=1\columnwidth]{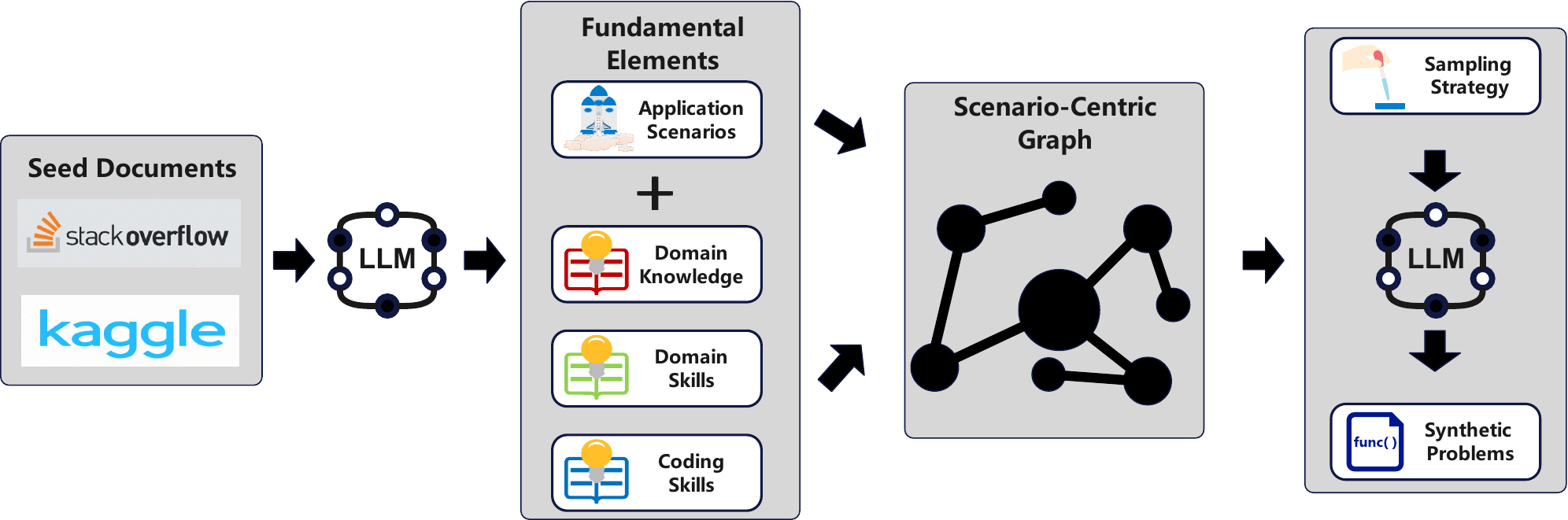}
    \caption{Illustration of the real-world programming problem synthesis framework}
    \label{fig:Illustration of the code problem synthesis framework}
\end{figure}

To address the limitations inherent in the lack of existing methodologies for synthesizing real-world programming tasks, we propose a novel framework, as illustrated in Figure~\ref{fig:Illustration of the code problem synthesis framework}. This framework begins with the curation of seed documents sourced from Stack Overflow and Kaggle, which effectively capture real-world coding tasks. From the documents, we extract four fundamental elements: application scenarios, domain knowledge, domain skills and coding skills. Based on these elements, we construct a scenario-centric knowledge graph, where the application scenario serves as the central node, connecting the corresponding knowledge and skill nodes. Subsequently, a sampling strategy is employed to select nodes from the knowledge graph, which are then utilized to prompt the LLM to synthesize high-quality real-world programming problems.

\subsection{Seed Curation}
\label{sec:seed_curation}

The foundation of our knowledge graph construction lies in the acquisition and rigorous processing of real-world programming resources. We curated seed data from two primary sources:
\begin{itemize}
    \item \textbf{Stack Overflow Dump}: programming discussions from the Stack Overflow dump
    \item \textbf{Kaggle Notebooks}: notebooks released by the Kaggle platform under an Apache 2.0 license
\end{itemize}

To ensure data quality for real-world problem synthesis, we implemented a multi-stage curation pipeline. The preprocessing phase applied strict filtering criteria: we retained only entries containing 500 to 20,000 characters, removed garbled text using regular expressions, and critically, preserved only Chinese and English content to maintain linguistic consistency. Deduplication proceeded through two complementary approaches: exact deduplication using hashing to remove identical entries, followed by near-duplicate detection applying MinHash algorithm. After these preprocessing steps, we obtained 11 million Stack Overflow entries and 1.5 million Kaggle notebooks. We then applied stratified subsampling to construct a balanced corpus, resulting in the final curated dataset of 3 million Stack Overflow entries and 0.5 million Kaggle notebooks.

\subsection{Foundamental Elements Extraction}
To the best of our knowledge, an effective coding problem or task is fundamentally designed to comprehensively assess an individual's proficiency across three interrelated dimensions: conceptual understanding of domain-specific knowledge, the application of domain-related skills, and the mastery of programming techniques. In light of this, we systematically extract and integrate these three dimensions from seed documents to synthesize a high-quality coding problem. Specifically, domain knowledge refers to a foundational concept or theoretical understanding pertinent to a particular field; domain skills encompass the practical application of methodologies and techniques within that domain knowledge, including their detailed implementation and contextual usage; and coding skills pertain to the adeptness in programming-related techniques and practices, with an emphasis on their precise application in real-world scenarios. Notably, coding skills can be further decomposed into three essential components: problem-solving and design thinking, tools and frameworks, and algorithms and data structures, all of which are critical for effectively addressing coding problems or completing programming development projects. 

In order to ensure that the synthesized coding problem is practical in nature and grounded in realistic application contexts, we also extract application scenarios from the seed documents. These application scenarios are characterized by their practical relevance and contextual diversity, enabling the generation of coding problems that closely mirror real-world development tasks. By leveraging these scenarios, we can effectively orchestrate the integration of domain knowledge, domain skills, and coding skills into a cohesive and meaningful problem. The detailed extraction prompt and illustrative example are provided in Appendix~\ref{appendix: Fundamental Elements Extraction Example}.

\subsection{Knowledge Graph Construction}
\label{subsec: Knowledge Graph Construction}
\begin{figure}
    \centering
    \includegraphics[width=1\columnwidth]{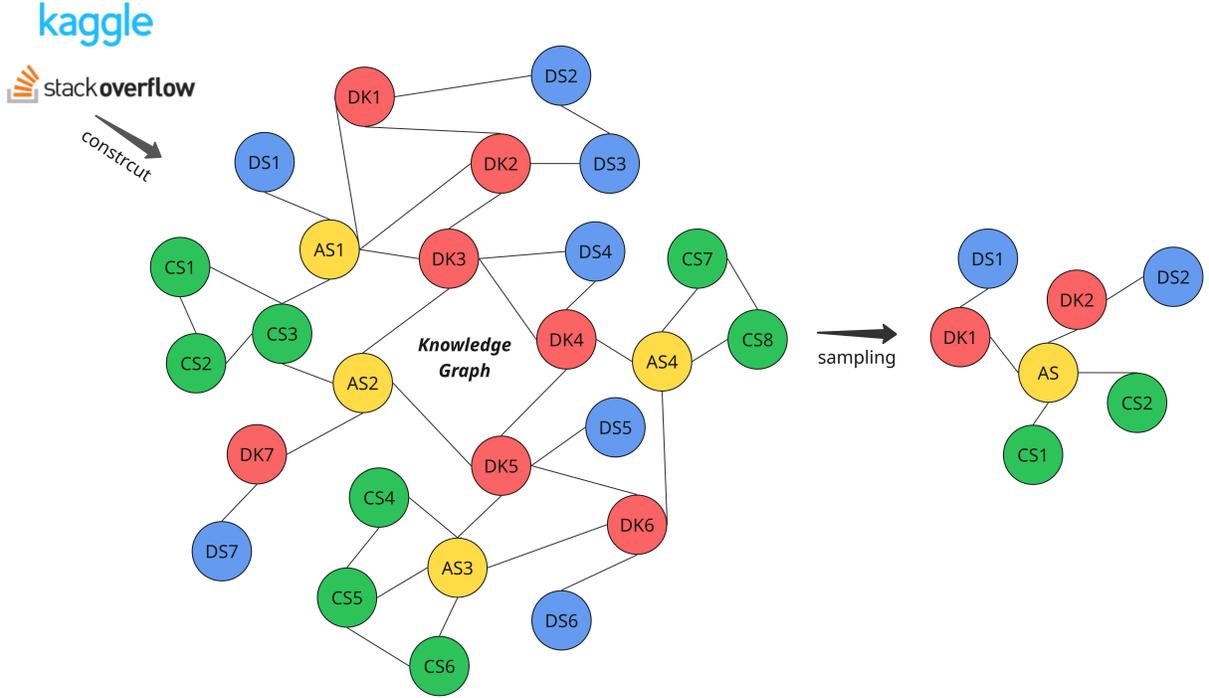}
    \caption{Scenario-centric Knowledge Graph: AS, DK, DS and CS denote application scenario, domain knowledge, domain skill and coding skill respectively.}
    \label{fig:kg}
\end{figure}

To construct a realistic problem and systematically model the interdependencies among conceptual and technical components in real-world programming tasks, we propose a scenario-centric knowledge graph (KG) centered around application scenarios. Each application scenario is linked to both domain knowledge and coding skills from the same seed data, thereby capturing the domain-specific knowledge and programming techniques pertinent to the scenario. Furthermore, we establish a connection between domain knowledge and domain skills in order to elucidate the practical application of the techniques and methodologies associated with that knowledge. Specifically, the KG, as illustrated in Figure~\ref{fig:kg}, is formally defined as $ G = (V, E) $, where $ V $ is the set of all entities (i.e., nodes) corresponding to the four defined elements, and $ E \subseteq V \times V $ is the set of edges representing the relationships between these entities.

Let $ \mathcal{AS} $, $ \mathcal{DK} $, $ \mathcal{DS} $, and $ \mathcal{CS} $ denote the sets of all application scenarios, domain knowledge, domain skills, and coding skills, respectively. The set of nodes $ V $ is then the union of these four sets:

$$
V = \mathcal{AS} \cup \mathcal{DK} \cup \mathcal{DS} \cup \mathcal{CS}.
$$

An edge $ (u, v) \in E $ exists if the corresponding entities $ u $ and $ v $ co-occur within the same document. The edge types are explicitly defined based on the types of nodes they connect. Specifically, the graph includes the following five types of relationships:

- $ (\mathcal{AS}, \mathcal{DK}) $: an application scenario is associated with relevant domain knowledge.

- $ (\mathcal{AS}, \mathcal{CS}) $: an application scenario requires certain coding skills.

- $ (\mathcal{DK}, \mathcal{DS}) $: domain knowledge is linked to the domain skills necessary for its application.

- $ (\mathcal{DK}, \mathcal{DK}) $: domain knowledge elements are connected if they co-occur in the same context.

- $ (\mathcal{CS}, \mathcal{CS}) $: coding skills are connected based on their co-occurrence.

This explicit typing of relationships is designed to support the generation of application scenario-driven programming problems. Starting from a given application scenario $ a \in \mathcal{AS} $, one can traverse the graph to identify the domain knowledge, domain skills, and coding skills required to construct a meaningful programming task. Furthermore, by allowing traversal within the subgraphs of domain knowledge and coding skills, the graph enables the inclusion of more nuanced and interrelated elements, thereby increasing the complexity and diversity of the generated problems.

\subsection{Sampling Strategy}
\label{subsec: Sampling Strategy}

At the core of our framework is the construction of a scenario-centric KG, which serves as the foundational structure for selecting an application scenario as the starting point in the generation of realistic programming problems. The integration of domain knowledge, domain skills, and coding skills connected with the chosen scenario forms the basis for problem synthesis, as shown in Figure~\ref{fig:kg}. This synthesis is achieved through a systematic sampling process that identifies interrelated and diverse combinations of these three elements. By stacking multiple interrelated combinations, the complexity of the generated problem is increased while maintaining internal coherence among the elements. Collectively, the scenario-centric KG, the diverse selection of nodes, and tunable complexity contribute to the effective construction of real-world programming problems.

\subsubsection{Transition Probabilities Among Nodes}

Following the construction of the KG, our objective is to leverage the KG to generate problems by selecting domain knowledge, domain skills, and coding skills connecting with the application scenario. To facilitate this process, we transform the co-occurrence frequencies between pairs of nodes into transition probabilities, as follows:

Let $ f(A, B) $ denote the frequency of co-occurrence between nodes $ A $ and $ B $. To derive a probabilistic traversal strategy, we first define the \textbf{first-step transition probability} from node $ A $ to its immediate connecting node $ B $ as:

{\fontsize{8}{0}\selectfont
$$
P^{(1)}(A \rightarrow B) = \frac{f(A, B)}{\sum_{B^\prime \in \mathcal{N}^{(1)}(A)} f(A, B^\prime)},
$$
}where $ \mathcal{N}^{(1)}(A) = \{ B^\prime \in V \mid (A, B^\prime) \in E \} $ is the set of all first-step neighbors of $ A $.

While first-step transitions are crucial for capturing direct knowledge realtionships, they maybe limited in terms of diversity. To expand the range of node selection and thereby generate more diverse problems, we extend the traversal mechanism to incorporate second-step neighbors, i.e., nodes that are two steps away from $ A $ through an intermediate node $ B^\prime $. Specifically, the set of second-step neighbors of $ A $, denoted $ \mathcal{N}^{(2)}(A) $, is defined as:

{\fontsize{8}{0}\selectfont
\begin{equation*}
\begin{split}
    \mathcal{N}^{(2)}(A) = \{ C \in V \mid C \notin \mathcal{N}^{(1)}(A) \text{ and } \exists B^\prime \in \mathcal{N}^{(1)}(A) \text{ such that } (B^\prime, C) \in E \}. 
\end{split}
\end{equation*}
}

The second-step transition probability from $ A $ to a second-step neighbor $ C $ is then defined as:

{\fontsize{8}{0}\selectfont
$$
P^{(2)}(A \rightarrow C) = \sum_{B^\prime \in \mathcal{N}^{(1)}(A)} P^{(1)}(A \rightarrow B^\prime) \cdot P^{(1)}(B^\prime \rightarrow C).
$$
}

By incorporating both first- and second-step transitions, we can not only capture direct relationship but also model more diverse, indirect relationships. To this end, we define the normalized combined transition probability of first- and second-step neighbors from node $ A $ to any node $ X \in \mathcal{N}^{(1)}(A) \cup \mathcal{N}^{(2)}(A) $ as:

{%
\fontsize{8}{0}%
\selectfont
\begin{flalign*}
    \begin{aligned}
  P_{\text{norm}}(A \rightarrow X) =
    \frac{P(A \rightarrow X)}{
      \sum_{X' \in \mathcal{N}^{(1)}(A)} P^{(1)}(A \rightarrow X')
      + \sum_{X'' \in \mathcal{N}^{(2)}(A)} P^{(2)}(A \rightarrow X'')}
    \end{aligned}
\end{flalign*}
}

Noted that second-step transitions often yield much smaller probabilities compared to first-step ones due to the compounded nature of the co-occurrence frequencies. To further enhance the diversity of the generated problems and to balance the influence of high- and low-probability transitions, we apply a softmax transformation with a temperature parameter $ T $ to the normalized transition probabilities.

The final transition probability is thus defined as:

{\fontsize{8}{0}\selectfont
$$
P(A \rightarrow Y) = \frac{ \exp\left( \frac{1}{T} \ln(P_{\text{norm}}(A \rightarrow Y)) \right) }{ \sum_{Y' \in \mathcal{N}^{(1)}(A) \cup \mathcal{N}^{(2)}(A)} \exp\left( \frac{1}{T} \ln(P_{\text{norm}}(A \rightarrow Y')) \right) },
$$
}where $ T = 1 $ corresponds to the original normalized distribution. As $ T $ increases, the distribution becomes flatter, giving higher weight to nodes with lower normalized probabilities and thereby promoting greater diversity and cross-domain coverage in the problem generation process.

\subsubsection{Random Sampling Strategy}

We initiate the process by randomly selecting an application scenario from the KG and traversing the respective edges $ (\mathcal{AS}, \mathcal{DK}) $, $ (\mathcal{DK}, \mathcal{DK}) $, $ (\mathcal{DK}, \mathcal{DS}) $, $ (\mathcal{AS}, \mathcal{CS}) $ and $ (\mathcal{CS}, \mathcal{CS}) $, guided by the transitional probabilities defined in the preceding subsection. Each traversal results in a combination of domain knowledge, domain skill, and coding skill, which we collectively refer to as a \textbf{feature}. This procedure is continued multiple times from the same initial application scenario to generate a set of interrelated features. Through this mechanism, we construct a programming problem whose level of complexity is systematically determined by the number of incorporated features. Accordingly, we define the \textbf{problem complexity} as the the number of features selected from the sampling process. The interdependencies among these features emerge naturally from the probabilistic structure of the scenario-centric KG, thereby reflecting the intricate and interconnected nature of real-world programming problems. This method, referred to as the random sampling strategy, enables the generation of a wide variety of problems with varying degrees of difficulty and domain coverage, enabling a more realistic and scalable problem generation framework.

\subsubsection{LLM-based Sampling Strategy}

Random sampling is fundamentally aligned with the intrinsic properties of the knowledge graph, as it preserves the natural distribution and interrelations among entities and relationships. An alternative approach involves the generation of a large candidate feature pool, followed by the application of an external mechanism, such as a LLM, to select a subset of the most interrelated features. Specifically, this method is structured into two sequential phases. First, ten candidate features are generated through the random sampling strategy, as outlined in the preceding subsection. Subsequently, Qwen3-32B with non-thinking mode is employed to select the most interrelated subset of these candidates, based on a predefined level of complexity. This allows for the creation of problems that are not only complex but also semantically coherent and logically interconnected. The prompt used in the LLM-based sampling process is detailed in appendix~\ref{ref: Prompt}.

\subsection{Problem Synthesis and Answer Generation}

Drawing upon the scenario-centric knoweldge graph, we employ the sampling strategy to select a predefined number of features, contingent upon the specified level of complexity. Utilizing these selected features, we leverage the Qwen3-32B model with non-thinking mode to generate real-world synthesis problems. The prompt is detailed in Appendix~\ref{ref: Prompt}. As the number of features increases, the complexity of the synthesis problems accordingly escalates. Representative examples of these problems are provided in Appendix~\ref{appendix: Samples of Synthesized Problems}.

For answer generation, we also employ the Qwen3-32B model with non-thinking mode to produce responses to synthesis questions. At this stage, we do not implement a dedicated mechanism for verifying the correctness of the generated solutions, as the primary objective of this study is to develop and evaluate the problem synthesis framework. While higher-quality responses can be achieved by utilizing a more powerful LLM or incorporating a sandbox environment for code validation, our experimental results demonstrate satisfactory performance even in the absence of such verification processes. We therefore defer the implementation of a formal answer verification mechanism to future work.

\section{Experiments}

\begin{table}[t]
    \centering
    \setlength{\tabcolsep}{1mm}
    \begin{tabular}{c|cc|cc|c|cc|cc}
    \toprule
    \multirow{3}{*}{\textbf{Model}} & \multicolumn{4}{c|}{\textbf{Real-world-level Benchmarks}} & \multicolumn{5}{c}{\textbf{Algorithm-level Benchmarks}}\\ 
    \cmidrule{2-10}
    & \multicolumn{2}{c|}{\textbf{BCB Instruct}} & \multicolumn{2}{c|}{\textbf{NCB (en)}} & \textbf{LCB} & \multicolumn{2}{c|}{\textbf{HumanEval}} & \multicolumn{2}{c}{\textbf{MBPP}}\\
    & \textit{Full} & \multicolumn{1}{c|}{\textit{Hard}} & \multicolumn{1}{c}{\textit{Python}} & \multicolumn{1}{c|}{\textit{Java}} & \textit{2408-2501} & \textit{HE} & \multicolumn{1}{c|}{\textit{HE+}} & \textit{MBPP} & \multicolumn{1}{c}{\textit{MBPP+}}\\ \midrule
    \multicolumn{10}{c}{$\sim 1.5\text{B Coder}$} \\ \midrule
    Deepseek-Coder-1.3B-Instruct & \multicolumn{1}{c}{21.75} & \multicolumn{1}{c|}{3.38} & 10.00 & \multicolumn{1}{c|}{24.30} & \multicolumn{1}{c|}{4.30} & 67.07 & \multicolumn{1}{c|}{62.20} & 63.23 & \multicolumn{1}{c}{53.44}\\
    Qwen2.5-Coder-1.5B-Instruct & \multicolumn{1}{c}{26.14} & \multicolumn{1}{c|}{3.38} & 18.60 & \multicolumn{1}{c|}{17.10} & \multicolumn{1}{c|}{6.93} & 65.85  & \multicolumn{1}{c|}{62.80} & 69.05 & \multicolumn{1}{c}{59.52}\\
    Qwen2.5-Coder-1.5B-SCoGen & \multicolumn{1}{c}{\textbf{32.46}} & \multicolumn{1}{c|}{\textbf{14.19}} & \textbf{20.00} & \multicolumn{1}{c|}{\textbf{28.60}} & \multicolumn{1}{c|}{\textbf{7.89}} & \textbf{68.29}  & \multicolumn{1}{c|}{\textbf{64.02}} & \textbf{73.81} & \multicolumn{1}{c}{\textbf{62.70}}\\
    \midrule
    \multicolumn{10}{c}{$\sim 7\text{B Coder}$} \\ \midrule
    OpenCoder-8B-Instruct & \multicolumn{1}{c}{40.53} & \multicolumn{1}{c|}{15.54} & 28.60 & \multicolumn{1}{c|}{24.30} & \multicolumn{1}{c|}{9.68} & 81.71 & \multicolumn{1}{c|}{77.44} & 81.75 & \multicolumn{1}{c}{71.43}\\
    CodeQwen1.5-7B-Chat & \multicolumn{1}{c}{37.37} & \multicolumn{1}{c|}{16.22} & 22.90 & \multicolumn{1}{c|}{32.90} & \multicolumn{1}{c|}{5.38} & 86.59 & \multicolumn{1}{c|}{80.49} & 80.95 & \multicolumn{1}{c}{69.84}\\
    CodeLlama-7B-Instruct & \multicolumn{1}{c}{21.84} & \multicolumn{1}{c|}{4.05} & 11.40 & \multicolumn{1}{c|}{17.10} & \multicolumn{1}{c|}{4.66} & 37.20 & \multicolumn{1}{c|}{31.71} & 53.44 & \multicolumn{1}{c}{43.92}\\
    Deepseek-Coder-6.7B-Instruct & \multicolumn{1}{c}{33.60} & \multicolumn{1}{c|}{10.81} & 24.30 & \multicolumn{1}{c|}{40.00} & \multicolumn{1}{c|}{10.75} & 81.10 & \multicolumn{1}{c|}{76.83} & 77.25 & \multicolumn{1}{c}{66.67}\\
    Qwen2.5-Coder-7B-Instruct & \multicolumn{1}{c}{41.05} & \multicolumn{1}{c|}{18.24} & 24.30 & \multicolumn{1}{c|}{32.90} & \multicolumn{1}{c|}{16.13} & \textbf{87.80} & \multicolumn{1}{c|}{\textbf{84.15}} & 84.39 & \multicolumn{1}{c}{71.96}\\
    Qwen2.5-Coder-7B-SCoGen & \multicolumn{1}{c}{\textbf{41.32}} & \multicolumn{1}{c|}{\textbf{24.32}} & \textbf{31.40} & \multicolumn{1}{c|}{\textbf{41.40}} & \multicolumn{1}{c|}{\textbf{19.00}} & 84.76 & \multicolumn{1}{c|}{83.54} & \textbf{85.71} & \multicolumn{1}{c}{\textbf{72.49}}\\
    \midrule
    \multicolumn{10}{c}{$\sim 7\text{B Model}$} \\ \midrule
    Qwen2.5-7B-Instruct & \multicolumn{1}{c}{36.75} & \multicolumn{1}{c|}{13.51} & 28.60 & \multicolumn{1}{c|}{21.40} & \multicolumn{1}{c|}{12.90} & \textbf{82.93} & \multicolumn{1}{c|}{75.61} & 79.10 & \multicolumn{1}{c}{67.99}\\
    Qwen2.5-7B-SCoGen & \multicolumn{1}{c}{37.37} & \multicolumn{1}{c|}{17.57} & \textbf{32.90} & \multicolumn{1}{c|}{32.90} & \multicolumn{1}{c|}{13.98} & 78.66  & \multicolumn{1}{c|}{74.39} & 79.10 & \multicolumn{1}{c}{66.40}\\
    Qwen3-8B & \multicolumn{1}{c}{40.96} & \multicolumn{1}{c|}{18.24} & 28.60 & \multicolumn{1}{c|}{\textbf{38.60}} & \multicolumn{1}{c|}{\textbf{25.45}} & 82.32 & \multicolumn{1}{c|}{\textbf{78.05}} & 81.48 & \multicolumn{1}{c}{\textbf{71.16}}\\
    Qwen3-8B-Base-SCoGen & \multicolumn{1}{c}{\textbf{42.19}} & \multicolumn{1}{c|}{\textbf{21.62}} & 30.00 & \multicolumn{1}{c|}{\textbf{38.60}} & \multicolumn{1}{c|}{22.22} & 79.21  & \multicolumn{1}{c|}{75.61} & \textbf{83.07} & \multicolumn{1}{c}{70.63}\\
    \bottomrule
    \end{tabular}
\caption{Main Result: Performance of models across multiple programming benchmarks. For our method, the parameters (random sampling, $ T = 3 $, $ C = 1 $) are selected based on the best performance reported in the Section~\ref{sec:ablation studies}.}
\label{tab:main result}
\end{table}

To evaluate the effectiveness of our proposed framework, we fine-tune the base models of code LLMs and compare them with their corresponding instruction-tuned counterparts. This comparison is designed to demonstrate the efficacy of our approach in enhancing model performance. Furthermore, we conduct supervised fine-tuning on the base version of a general-purpose LLM to investigate whether our framework can also improve the code generation capabilities of models not specifically trained for coding tasks.

\subsection{Setup}

\subsubsection{Training Setup}

We conduct supervise fine-tunning (SFT) on Qwen2.5-Coder-1.5B \citep{hui2024qwen25codertechnicalreport}, Qwen2.5-Coder-7B \citep{hui2024qwen25codertechnicalreport}, Qwen2.5-7B \citep{qwen2025qwen25technicalreport} models and Qwen3-8B-Base \citep{yang2025qwen3technicalreport}, utilizing a dataset of 500K synthesized real-world programming question-and-answer pairs. The detailed training configurations is in Appendix~\ref{ref:Training Details}.

\subsubsection{Evaluation Benchmarks}

We conduct our evaluation across two distinct benchmark categories: real-world-level tasks and algorithm-level tasks. For the real-world-level evaluation, we utilize the BigCodeBench Instruct (BCB Instruct) \citep{zhuo2024bigcodebench} and the NaturalCodeBench English (NCB (en)) \citep{zhang-etal-2024-naturalcodebench} benchmarks, which are representative of real-world coding scenarios in natural language settings. To assess performance on challenging algorithm problems, we employ the LiveCodeBench (LCB) \citep{jain2024livecodebench} from 2408 to 2501. In addition to LCB, we further evaluate our model on widely used single-function code generation benchmarks, including HumanEval \citep{chen2021codex}, HumanEval+ \citep{evalplus}, MBPP \citep{austin2021program}, and MBPP+ \citep{austin2021program}. These benchmarks collectively allow us to comprehensively assess the model's code generation capabilities across diverse and complex programming tasks. A detailed description of the benchmark datasets is provided in Appendix~\ref{ref:Benchmarks}.

\subsubsection{Baselines}

We conduct a comprehensive comparison with open-source state-of-the-art models in the size of approximately 1.5B and 7B. The detailed baseline models and inference setup is in Appendix~\ref{ref:Baseline Details}.

\subsection{Main Results}

As illustrated in Table~\ref{tab:main result}, our fine-tuned model demonstrates substantial improvements in addressing real-world programming tasks. Specifically, our fine-tuned Qwen2.5-Coder-7B achieves an accuracy of 24.32\% on the BigCodeBench Instruct Hard, 31.40\% (Python) and 41.40\% (Java) on the NaturalCodeBench. These results represent absolute improvements of 6.08\%, 7.10\%, and 8.50\%, respectively, over the Qwen2.5-Coder-7B-Instruct model. A similar performance enhancement is observed across our fine-tuned 1.5B coder and general 7B models, indicating that our synthetic data generation approach effectively boosts model performance during SFT on practical programming tasks. In addition, our models demonstrate superior performance compared to all other shown open-souce state-of-art models.

In addition to real-world tasks, our model also exhibits strong performance on algorithm-level benchmarks. On the LiveCodeBench benchmark, our fine-tuned Qwen2.5-Coder-7B achieves an accuracy of 19.00\%, significantly outperforming the Qwen2.5-Coder-7B-Instruct, which attains 16.13\%. Moreover, on HumanEval and MBPP, our model achieves performance that is comparable to the baseline.

This consistent improvement is not limited to a single model variant but is also observed across our fine-tuned models with varying sizes and architectural configurations. Notably, the fine-tuned Qwen3-8B-Base model demonstrates competitive performance. On real-world-level benchmarks, it outperforms the Qwen3-8B model by an average of 1.5\%, and on algorithm-level benchmarks, it achieves results within 1.5\% of the Qwen3-8B's performance on average, using only SFT. It is worth noting that the Qwen3-8B model undergoes a distillation process, which results in better performance than the combination of SFT and RL. This process enables the student model to expand its exploration space and enhance its reasoning potential \citep{yang2025qwen3technicalreport}. Importantly, our fine-tuned model is able to achieve comparable performance using SFT alone.

These results suggest that our framework synthesizes high-quality questions that not only improve the model’s capability in solving complex, real-world tasks but also maintain competitive performance on algorithm-level problems.

\subsection{Ablation Studies}
\label{sec:ablation studies}

To systematically evaluate the influence of individual components on the synthesis process, we conducted ablation studies focusing on three key factors: sampling strategy, problem complexity, and the temperature parameter.

Regarding the sampling strategy, we compare two approaches: random sampling and the LLM-based sampling strategy. These methods are evaluated to determine their relative effectiveness in selecting the interrelated features.

In the context of problem complexity, we investigate three predefined levels to assess their impact on the performance of SFT. Specifically, complexity level C1 involves a single feature comprising three elements (a domain knowledge, domain skill, and coding skill), C2 includes two features with a total of six elements, and C3 encompasses three features with nine elements. This hierarchical design allows us to examine how increasing the complexity of the training data affects the SFT outcomes.

The temperature parameter controls the distribution of transition probabilities in the synthesis process. In this study, we investigate the effects of three distinct temperature values: $ T = 1 $ (T1), $ T = 2 $ (T2), and $ T = 3 $ (T3), in order to assess the impact of diversity on the outcomes of SFT.

For all configurations, we conduct the SFT on Qwen2.5-Coder-7B, and the size of the training dataset is 500K and real-world-level benchmarks are employed to ensure a fair and controlled comparison across different settings.

\subsubsection{Sampling Strategy}

\begin{table}[t]
    \centering
    \setlength{\tabcolsep}{1mm}
    \begin{tabular}{l|cc|cc|c}
    \toprule
    \multirow{2}{*}{\textbf{Model}} & \multicolumn{2}{c|}{\textbf{BCB Instruct}} & \multicolumn{2}{c|}{\textbf{NCB (en)}}& \multirow{2}{*}{\textbf{Average}}\\
     & \textit{\ Full} & \multicolumn{1}{c|}{\textit{Hard}} & \multicolumn{1}{c}{\textit{Python}} & \multicolumn{1}{c|}{\textit{Java}}\\ \midrule
    Random & \ 41.40 & \multicolumn{1}{c|}{22.75}  & \multicolumn{1}{c}{31.28} & \multicolumn{1}{c|}{38.23} & \multicolumn{1}{c}{\textbf{33.42}}\\
    LLM-based & \ 41.44 & \multicolumn{1}{c|}{22.52} & \multicolumn{1}{c}{32.09} & \multicolumn{1}{c|}{33.49} & \multicolumn{1}{c}{32.39}\\
    \bottomrule
    \end{tabular}
\caption{Ablation Studies on Sampling Strategy: Average performance of models with two sampling strategies on multiple programming benchmarks}
\label{tab:ablation studies: sampling strategy}
\end{table} 

The ablation study on the sampling strategy is summarized in Table~\ref{tab:ablation studies: sampling strategy}. The reported accuracy values represent the average performance across all nine configurations, which are combinations of the three complexity settings (C1, C2, C3) and the three temperature parameters (T1, T2, T3) for each sampling strategy. The random sampling strategy achieves an average accuracy of 33.42\%, which is higher than the 32.39\% obtained using the LLM-based sampling approach. These results indicate that the scenario-centric graph, in conjunction with the transition probability mechanism, is sufficiently effective in generating real-world programming tasks without external guidance from LLMs. Furthermore, we posit that relying on LLMs to select interrelated features may introduce a selection bias. Consequently, this could lead to a reduction in the diversity of the synthesized programming questions.

\subsubsection{Complexity and Temperature}

\begin{figure}
    \centering
    \includegraphics[width=1\linewidth]{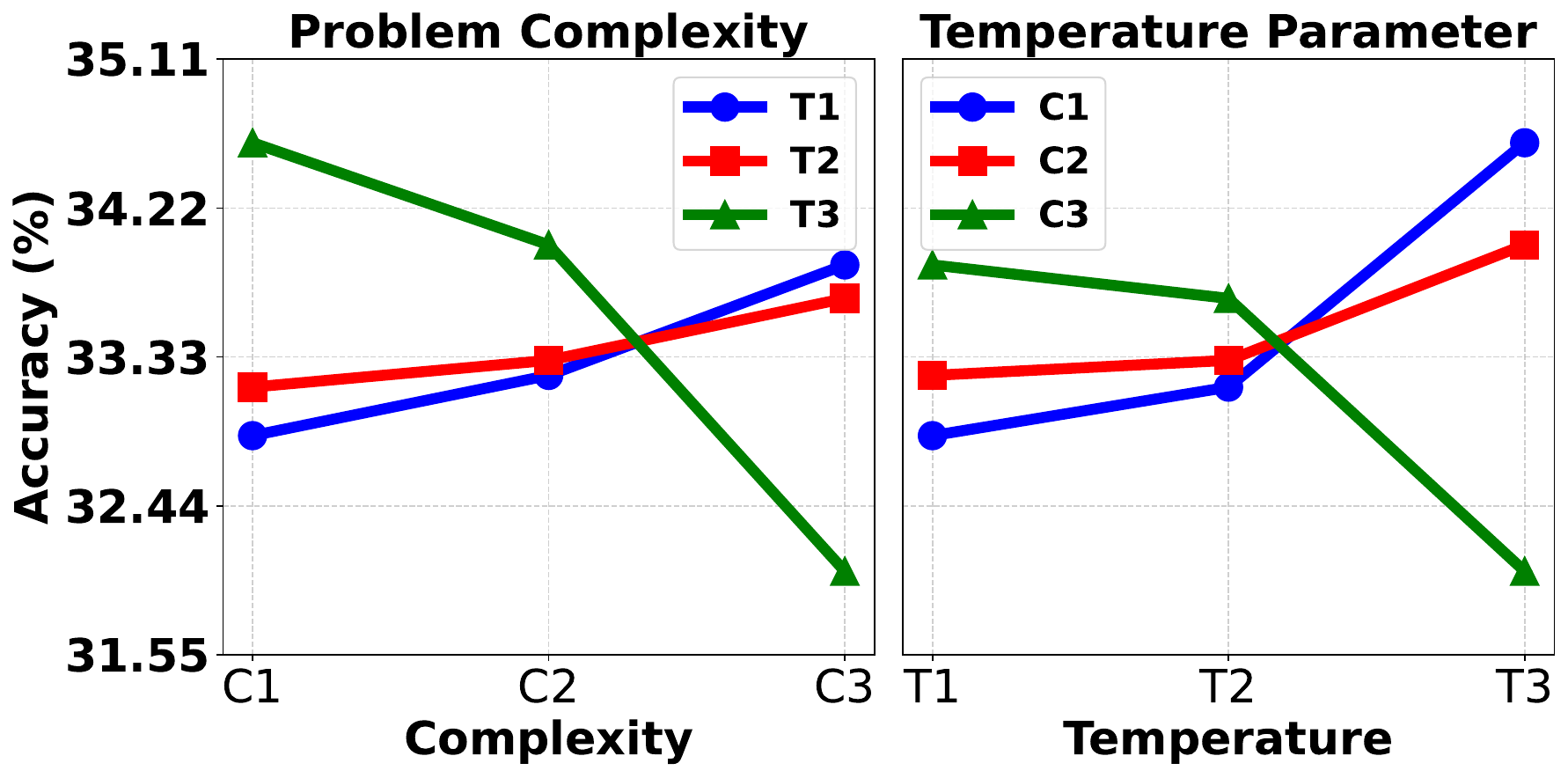}
    \caption{Abaltion Studies: Performance of models with different complexity and temperature}
    \label{fig:ablation studies}
\end{figure}

The detailed results of the nine configurations under the random sampling strategy are presented in Table~\ref{tab:ablation studies}, while the corresponding line plot is displayed in Figure~\ref{fig:ablation studies}. The experimental findings reveal a general trend: as the problem complexity increases, the accuracy exhibits a consistent upward trajectory under a fixed temperature setting. Specifically, under temperature settings T1 and T2, as illustrated on the left-hand side of Figure~\ref{fig:ablation studies}, the accuracy increases with the rise in problem complexity. Moreover, a general trend is observed wherein accuracy tends to improve with an increase in temperature, assuming a constant level of complexity, as demonstrated on the right-hand side of Figure~\ref{fig:ablation studies}.

However, an exception is observed under the combination of complexity C3 and temperature setting T3, where the model's accuracy exhibits a declining trend in response to the simultaneous increase in temperature and problem complexity. This phenomenon can be attributed to the composition of C3, which incorporates three distinct features—representing three domain knowledge, three domain skills, and three coding skills respectively—resulting in a total of nine elements. Under high temperature settings, the diversity of these elements is further amplified. Consequently, the synthesized questions may consist of knowledge and skill components that lack coherence, leading to a deviation from realistic and meaningful problem scenarios. This observation implies that caution should be exercised when generating questions that incorporate a large number of highly diverse elements. It is therefore essential to strike a careful balance between the diversity, complexity, and coherence of the synthesized problems.

It is noteworthy that the results of all nine models presented in Table~\ref{tab:ablation studies} outperform those of the Qwen2.5-Coder-7B-Instruct baseline. Based on these findings, we provide the following recommendations regarding the model configuration settings:

\begin{itemize}
    \item The C1T1 setting is not recommended due to its limited diversity in the generated outputs.
    \item Conversely, the C3T3 setting introduces excessive diversity, which may lead to a degradation in performance.
    \item Intermediate configurations, however, demonstrate a balanced trade-off between diversity and performance, yielding satisfactory results.
\end{itemize}

Therefore, it is advisable to adopt model configurations that lie between C1T1 and C3T3 for optimal performance and diversity.

\begin{table}[t]
    \centering
    \begin{tabular}{l|cc|cc|c}
    \toprule
    \multirow{2}{*}{\textbf{Model}} & \multicolumn{2}{c|}{\textbf{BCB Instruct}} & \multicolumn{2}{c|}{\textbf{NCB (en)}}& \multirow{2}{*}{\textbf{Average}}\\
     & \textit{\ Full} & \multicolumn{1}{c|}{\textit{Hard}} & \multicolumn{1}{c}{\textit{Python}} & \multicolumn{1}{c|}{\textit{Java}}\\ \midrule
    C1T1 & \ 41.32 & \multicolumn{1}{c|}{23.00}  & \multicolumn{1}{c}{30.00} & \multicolumn{1}{c|}{37.10} & \multicolumn{1}{c}{32.86}\\
    C2T1 & \ 41.40 & \multicolumn{1}{c|}{22.97} & \multicolumn{1}{c}{31.40} & \multicolumn{1}{c|}{37.10} & \multicolumn{1}{c}{33.22}\\
    C3T1 & \ 40.26 & \multicolumn{1}{c|}{20.95} & \multicolumn{1}{c}{32.90} & \multicolumn{1}{c|}{41.40} & \multicolumn{1}{c}{33.88}\\
    C1T2 & \ 41.75 & \multicolumn{1}{c|}{23.65} & \multicolumn{1}{c}{28.60} & \multicolumn{1}{c|}{38.60} & \multicolumn{1}{c}{33.15}\\
    C2T2 & \ 41.75 & \multicolumn{1}{c|}{22.97} & \multicolumn{1}{c}{31.40} & \multicolumn{1}{c|}{37.10} & \multicolumn{1}{c}{33.31}\\
    C3T2 & \ 40.88 & \multicolumn{1}{c|}{20.95} & \multicolumn{1}{c}{34.30} & \multicolumn{1}{c|}{38.60} & \multicolumn{1}{c}{33.68}\\
    C1T3 & \ 41.32 & \multicolumn{1}{c|}{24.32} & \multicolumn{1}{c}{31.40} & \multicolumn{1}{c|}{41.40} & \multicolumn{1}{c}{\textbf{34.61}}\\
    C2T3 & \ 41.67 & \multicolumn{1}{c|}{24.32} & \multicolumn{1}{c}{32.90} & \multicolumn{1}{c|}{37.10} & \multicolumn{1}{c}{34.00}\\
    C3T3 & \ 42.28 & \multicolumn{1}{c|}{21.62} & \multicolumn{1}{c}{28.60} & \multicolumn{1}{c|}{35.70} & \multicolumn{1}{c}{32.05}\\
    \bottomrule
    \end{tabular}
\caption{Ablation Studies on Random Sampling Strategy: Performance of models with various configuration on real-world-level programming benchmarks}
\label{tab:ablation studies}
\end{table}

\section{Conclusion and Future Works}

In this paper, we address the challenge of the scarcity of real-world programming problems by proposing a novel framework for synthesizing such problems. Our framework systematically extracts domain knowledge, domain skills, and coding skills from a curated set of seed documents, including Stack Overflow and Kaggle. These extracted elements serve as the foundational components for constructing programming questions. In order to formulate problems that reflect realistic settings and practical applications, we extract application scenarios from the same seed documents. These application scenarios act as central nodes in the scenario-centric graph, interconnecting all related knowledge and skill nodes. A sampling strategy is then applied to select interrelated features, enabling the generation of diverse and complex real-world programming problems. Experimental results demonstrate that our framework outperforms open-source state-of-the-art large language models on real-world-level and algorithm-level benchmarks. In future work, we intend to explore an answer verification mechanism to further enhance the robustness of our approach. Additionally, we aim to investigate the scalability of our framework by applying it to larger models with 32B or more.

\section{Limitations}

Although our framework can synthesize real-world problems involving multiple domains, multiple programming languages, and the generation of multiple functions and files, they do not yet fully encompass repository-level challenges. The current set of questions is predominantly focused on code generation tasks. To advance this research direction, further investigation is required to develop high-quality repository-level questions and benchmark datasets.

\bibliographystyle{plainnat}
\bibliography{reference}

\appendix

\section{Real-world vs Algorithm-level}
\label{appendix: Real-world vs Algorithm-level}
\begin{figure}[H]
    \centering
    \includegraphics[width=1\linewidth]{function_compare.pdf}
    \caption{Real-world vs. algorithm-level code implementation}
    \label{fig:intro 1}
\end{figure}

\section{Prompt}
\label{ref: Prompt}

This section presents the prompts utilized in the Methodology. These include the prompts employed for the fundamental elements extraction, the LLM-based sampling strategy, and the problem synthesis.

\begin{tcolorbox}[breakable, title=Prompt: Fundamental Elements Extraction]
You are a code-related text analysis expert. Given a piece of code-related text, you will extract the following attributes:\\
\\
1. **Application Scenario**: Extract the most specific, concrete real-world application scenario where this code/algorithm would be practically used.\\
\\
**Guidelines for Application Scenario**:\\
- Focus on WHERE and HOW this code would be used in real software systems\\
- Avoid generic categories like "data processing", "mathematical computation", "algorithm implementation"\\
- Think about specific industries, use cases, or problem domains\\
- Consider what kind of software system or application would need this functionality\\
- If multiple scenarios are possible, choose the most common or practical one\\
\\
**Examples of good vs bad scenario extraction**:\\
Bad: "Mathematical Computation Tool", "Data Processing System", "Algorithm Implementation"  
Good: "Computer Graphics Engine Curve Rendering", "Financial Trading Platform Risk Calculation Module", "Scientific Computing Software Symbolic Math Engine"\\
\\
2. **Domain Knowledge**:  \\
Identify 1 to 3 key domain concepts and its usage of this knowledge (in less than 15 words) that are most relevant and thoroughly discussed in the text.  \\
The concepts and usage should be detailed and specific, but expressed in general terms without reference to problem-specific details. The concepts may come from different domains.\\
Format: Domain Knowledge: Detail Usage  \\
(e.g., "XGBoost Regression: Predict target variable using gradient boosting decision trees with hyperparameter tuning.", 
"ARIMA Modeling: Fit and forecast time series data using autoregressive integrated moving average models.")
\\
3. **Domain Skill**: For each domain knowledge, extract up to one associated skill/method and the usage (if exists) that 
represents a problem-solving technique related to that knowledge.\\
The skill/method should be directly related to the concept and *applied in the provided text*.  \\
- **If no clear skill is present, write "NA".**\\
- Avoid forcing the extraction of a skill if the text does not *deeply* involve one. \\
- If the technique is too subtle, write "NA".\\
- Provide a concise, detailed explanation of the skill in general terms. \\ 
Format: Domain Skill: Detail methods to achieve it.\\
(e.g., "Elbow Method: Determine the optimal number of clusters by analyzing variance explained versus number of clusters.",
"AutoARIMA: Automatically select optimal ARIMA parameters using statistical criteria and grid search techniques.")
\\
4. **Coding Skill**: Extract *one* core programming logic for each category: problem-solving and design thinking, 
tools and frameworks, as well as algorithms and data structures\\
\\
(e.g., "Algorithms and Data Structures
1. Data Querying and Aggregation Analysis: Perform statistical, filtering, and aggregation operations on air quality data through SQL queries.")\\
\\
- If the text doesn’t involve coding or the information is not present, output "NA".\\
\\
\#\#**Note:**\\
Do not output any explanation, output only as the format below.\\
\\
\#\# Output Format (output in English)\\
\{output\_format\}\\
\\
\#\#**Code Text**\\
\{code\_text\}\\
\\
\#\# Output
\end{tcolorbox}

\begin{tcolorbox}[breakable, title=Prompt: LLM-based Sampling Strategy]
    
You will be provided with three groups of feature descriptions, with 10 items in each group. These features are essential elements for constructing a complex coding problem in a real-world scenario. Your task is to deeply understand the meaning of these features and their usage strategies in real coding scenarios, and then select \texttt{\{number\}} most appropriate elements, \texttt{\{number\}} from each group, such that the selected elements can generate a single, natural and realistic coding problem.

Feature descriptions include the following:
\begin{itemize}
    \item Domain Knowledge: A specific piece of knowledge or understanding relevant to the field.
    \item Domain Skill: A specific skill or method used in the domain, along with its detailed usage.
    \item Coding Skill: A specific programming-related skill or technique, along with its detailed usage.
\end{itemize}

Guidelines:
\begin{itemize}
    \item The domain knowledge and domain skill have already been paired; please do not separate them.
    \item The selected elements need to play distinct roles, thereby naturally leading to a complex question with a unified problem context.
    \item The selected elements should balance relevance and diversity.
\end{itemize}

Feature Descriptions:\\
\{DK\_DS\_features\}\\

\{CS\_features\}\\

First provide a concise step-by-step thought process, then give the selected elements (only the label is needed) as the following format:\\
\\
Step-by-Step Thought Process\\
\\
Selected Elements:\\
\\
\{output\_format\}

\end{tcolorbox}

\begin{tcolorbox}[breakable, title=Prompt: Problem Synthesis]
\textbf{You are a problem designer.} I will provide you with one or more features. Based on these, your task is to create a \textbf{single, cohesive real world coding problem} that integrates the provided features into a natural and practical context.

Each feature will include the following three traits:

\begin{itemize}
    \item \textbf{Domain Knowledge}: A specific piece of knowledge or understanding relevant to the field.
    \item \textbf{Domain Skill}: A specific skill or method used in the domain, along with its detailed usage.
    \item \textbf{Coding Skill}: A specific programming-related skill or technique, along with its detailed usage.
\end{itemize}

\textbf{Important Guidelines:}
\begin{itemize}
    \item First, identify a suitable real-world application scenario based on the given features. Then, develop a detailed programming problem of that scenario, ensuring it aligns with the features.
    \item \textbf{Do not mention the domain or coding skills explicitly} in the problem statement. Instead, \textbf{design the scenario in such a way that the solution naturally involves applying those skills}.
    \item The problem should be a \textbf{single, substantial task}, not a list of subtasks. The features should be \textbf{interconnected}, with one depending on or influencing another.
    \item If there is a \textbf{conflict between the provided features}, \textbf{resolve the conflict} by using only the most relevant or compatible parts of the features.
    \item The final output should be a \textbf{realistic, natural, and technically sound coding problem} that reflects a real-world scenario and integrates the given features in a meaningful way.
    \item If the question require the usage of datasets, provide schema and examples of the dataset.
    \item Do not generate any bonus or optional challenges.
\end{itemize}

\textbf{Features:}\\
\{features\}

First provide a concise step-by-step thought process, then generate the real world coding problem:\\

\textbf{Output Format:}\\
\{output\_format\}

\end{tcolorbox}

\section{Fundamental Elements Extraction Example}
\label{appendix: Fundamental Elements Extraction Example}

In this section, we present an example of the four elements extracted from a seed document: application scenario, domain knowledge, domain skills, and coding skills. For domain knowledge, domain skills, and coding skills, we extract both the knowledge or skill itself and its detailed usage within the seed document, formatted as \texttt{knowledge/skill: detailed usage}.

When constructing a scenario-centric knowledge graph (for detailed methodology, refer to Subsection~\ref{subsec: Knowledge Graph Construction}), only the knowledge or skill component is utilized as a node to establish connections. The detailed usage, on the other hand, serves as a mechanism to enhance the diversity of the synthesized problems following the sampling process (for further details, see Subsection~\ref{subsec: Sampling Strategy}). Specifically, for each element selected from the interconnected features, a randomly chosen corresponding detailed usage is appended to the associated knowledge or skill node.

\begin{tcolorbox}[breakable, title=Fundamental Elements Extraction Example]

\textbf{Application Scenario:}  
Medical Imaging Diagnostic System for Breast Cancer Detection

\textbf{Domain Knowledge:}  
\begin{enumerate}
    \item[1.] PyTorch Deep Learning Framework: Build and train neural networks for medical image classification
    \item[2.] DICOM Image Processing: Read and normalize medical imaging data for model input
    \item[3.] Stratified Sampling: Ensure balanced representation of classes in training and validation sets
\end{enumerate}

\textbf{Domain Skill:}  
\begin{enumerate}
    \item[1.] PyTorch Deep Learning Framework:
        \begin{enumerate}[labelindent=10pt, leftmargin=*, labelwidth=1.2cm]
            \item[1.1.] Transfer Learning: Fine-tune pre-trained convolutional neural networks for medical image classification
        \end{enumerate}
    \item[2.] DICOM Image Processing:
        \begin{enumerate}[labelindent=10pt, leftmargin=*, labelwidth=1.2cm]
            \item[2.1.] Pixel Normalization: Scale pixel values for consistent input to deep learning models
        \end{enumerate}
    \item[3.] Stratified Sampling:
        \begin{enumerate}[labelindent=10pt, leftmargin=*, labelwidth=1.2cm]
            \item[3.1.] Stratified K-Fold Cross Validation: Partition dataset while preserving class distribution for reliable model evaluation
        \end{enumerate}
\end{enumerate}

\textbf{Coding Skill:}  \\
\textbf{Problem-solving and Design Thinking:}
\begin{enumerate}
    \item[1.] Medical Image Preprocessing Pipeline: Normalize and visualize DICOM images for model training
\end{enumerate}

\textbf{Tools and Frameworks:}  
\begin{enumerate}
    \item[1.] PyTorch and Scikit-learn Integration: Combine deep learning with traditional ML tools for data handling and evaluation
\end{enumerate}

\textbf{Algorithms and Data Structures:}  
\begin{enumerate}
    \item[1.] Pixel Array Manipulation: Apply mathematical transformations to medical imaging data for visualization and preprocessing
\end{enumerate}

\end{tcolorbox}

\section{Samples of Synthesized Problems}
\label{appendix: Samples of Synthesized Problems}

In this section, we present three synthesized real-world programming problems, each characterized by a distinct level of complexity. All examples are generated with a transition probability temperature parameter set to 2. Complexity 1 involves the integration of a single feature---comprising one domain knowledge, one domain skill, and one coding skill. Complexity 2 incorporates two such interrelated features, while complexity 3 includes three interrelated features in the problem synthesis process.

\subsection{Complexity 1 Temperature 2}

\begin{tcolorbox}[breakable, title=Synthetic Real-world Programming Question]

You are working on a large enterprise software deployment system that integrates multiple third-party libraries. Occasionally, some assemblies fail to bind during runtime, and these failures are logged in a structured format for diagnostic purposes.

You have been provided with a dataset in CSV format that captures the details of these binding failures. Each log entry contains the name of the assembly that failed to bind, the time of the failure, the machine ID where it occurred, and a diagnostic message.

Your task is to analyze this dataset and generate a report that summarizes the \textbf{top 5 most frequently failing assemblies}, along with the \textbf{number of failures per assembly} and the \textbf{set of unique machine IDs} where those failures occurred.

\textbf{Dataset Schema:}
\begin{itemize}
    \item \texttt{timestamp}: datetime (e.g., 2025-03-20 14:22:33)
    \item \texttt{assembly\_name}: string (e.g., "MyLibrary.dll")
    \item \texttt{machine\_id}: string (e.g., "MACHINE-12345")
    \item \texttt{diagnostic\_message}: string (e.g., "Failed to resolve assembly version 1.2.3.4")
\end{itemize}

\textbf{Output Requirements:}
\begin{itemize}
    \item A list of the \textbf{top 5 assemblies} with the \textbf{most failures}, sorted in descending order of failure count.
    \item For each of these assemblies, include:
    \begin{itemize}
        \item The \textbf{number of failures}
        \item The \textbf{set of unique machine IDs} where the failures occurred
    \end{itemize}
\end{itemize}

\textbf{Input:}
\begin{itemize}
    \item A CSV file named \texttt{assembly\_logs.csv} with the schema described above.
\end{itemize}

\textbf{Output:}
\begin{itemize}
    \item A single Python script that reads the CSV file and prints the summary as described.
\end{itemize}

\textbf{Notes:}
\begin{itemize}
    \item You may assume the dataset is large and requires efficient processing.
    \item You should avoid printing unnecessary data or intermediate steps.
    \item Ensure your script handles any potential missing or malformed data gracefully.
\end{itemize}

\end{tcolorbox}

\subsection{Complexity 2 Temperature 2}

\begin{tcolorbox}[breakable, title=Synthetic Real-world Programming Question]
You are tasked with building a system to monitor and visualize stock price trends over time. The system consists of two parts:

\textbf{1. Backend Service}
The backend service that:
\begin{itemize}
    \item Fetches historical stock price data from a public financial API (e.g., Alpha Vantage or Yahoo Finance).
    \item Stores this data in a MySQL database for long-term persistence.
    \item Is configured using a build tool (e.g., Maven), and includes all necessary dependencies for API calls and database interaction.
\end{itemize}

\textbf{2. iOS App}
The iOS app that:
\begin{itemize}
    \item Displays the stock price data from the backend.
    \item Allows the user to select a stock symbol and time range.
    \item Passes the selected data to a new view controller that visualizes the stock trend as a line chart.
\end{itemize}

\textbf{Requirements}
\begin{itemize}
    \item The backend must be set up to run as a system service (e.g., MySQL is enabled and started at boot).
    \item The backend must resolve all necessary dependencies (e.g., libraries for HTTP requests and database access) using a dependency management file.
    \item The iOS app must pass the selected stock symbol and time range to the next view controller using appropriate lifecycle methods.
    \item The iOS app must make an API call to the backend to retrieve the data and render it in a chart.
\end{itemize}

\textbf{Dataset Example}

\textbf{Stock Data (JSON format from API):}
\begin{lstlisting}[language=json,backgroundcolor=\color{gray!10}]
{
    "Meta Data": {
    "symbol": "AAPL",
    "function": "TIME_SERIES_DAILY",
    "interval": "1day",
    "outputsize": "10",
    "time_zone": "UTC"
    },
    "Time Series (Daily)": {
    "2024-04-05": {"1. open": "190.00000", "4. close": "191.00000"},
    "2024-04-04": {"1. open": "189.00000", "4. close": "189.50000"},
    ...
    }
}
\end{lstlisting}

\textbf{Database Schema (MySQL):}
\begin{lstlisting}[language=SQL,backgroundcolor=\color{gray!10}]
CREATE TABLE stock_data (
    id INT AUTO_INCREMENT PRIMARY KEY,
    symbol VARCHAR(10),
    date DATE,
    open_price DECIMAL(10,2),
    close_price DECIMAL(10,2)
);
\end{lstlisting}

Your task is to implement the full system described above, ensuring that all components are correctly configured, communicate effectively, and operate as a cohesive solution for stock price monitoring and visualization.

\end{tcolorbox}

\subsection{Complexity 3 Temperature 2}

\begin{tcolorbox}[breakable, title=Synthetic Real-world Programming Question]
    You are tasked with building a diagnostic prediction system for a rare neurological condition using multimodal patient data. The system must process 3D brain scans, clinical text reports, and longitudinal patient health records to predict the likelihood of the condition.

    \textbf{Dataset Description:}
    
    You are provided with a dataset containing the following files for each patient:
    
    \begin{itemize}
        \item \texttt{brain\_scan.nii.gz}: A 3D NIfTI file containing the brain scan.
        \item \texttt{clinical\_report.txt}: A clinical report written by a neurologist, describing symptoms and findings.
        \item \texttt{health\_record.csv}: A time-series CSV with patient health metrics over time (e.g., blood pressure, heart rate, etc.).
        \item \texttt{segmentation\_rle.txt}: A run-length encoded string representing a binary segmentation mask of a suspicious brain region in the NIfTI scan.
        \item \texttt{label.txt}: A binary label indicating whether the patient has the condition (1) or not (0).
    \end{itemize}
    
    \textbf{Problem Statement:}
    
    Your task is to implement a complete diagnostic pipeline that:
    
    \begin{enumerate}
        \item Loads and preprocesses the NIfTI scan, decoding the run-length encoded mask and aligning it to the 3D image.
        \item Extracts visual features from the brain scan using a pre-trained Vision Transformer and combines them with textual features from the clinical report using a pre-trained BERT model.
        \item Processes the patient's health time-series data using an LSTM model to capture temporal patterns.
        \item Uses the outputs of the three models (image-text fusion, LSTM) as base learners and combines them using a SuperLearner meta-learner to produce the final diagnostic prediction.
    \end{enumerate}
    
    Your implementation must handle variable-length sequences in both the health time-series and the clinical text, and must be efficient for batch processing of variable-sized 3D scans.
    
    You are expected to write code that can be deployed to process a new batch of unlabeled patient data, producing a prediction for each patient.
    
    \textbf{Deliverables:}
    
    \begin{itemize}
        \item A complete Python script that implements the diagnostic pipeline.
        \item Proper use of PyTorch's \texttt{DataLoader} and a custom collate function to handle variable-length sequences.
        \item Integration of multimodal data (3D imaging, text, and time-series) into a single prediction model.
        \item Use of a meta-learner to combine model outputs for final prediction.
    \end{itemize}
    
    \section{Notes:}
    
    \begin{itemize}
        \item You may assume pre-trained models are available for use.
        \item You may use libraries such as \texttt{nibabel}, \texttt{transformers}, and \texttt{torch} as needed.
        \item Ensure your code is modular and well-documented for deployment.
    \end{itemize}
\end{tcolorbox}

\section{Training Details}
\label{ref:Training Details}
The training process of Qwen2.5-Coder-1.5B, Qwen2.5-Coder-
7B, Qwen2.5-7B and Qwen3-8B-Base employs full-parameter fine-tuning and consumes 213, 265, 317 and 451 Ascend 910B4 NPU hours respectively. The hyperparameters for training are configured as follows: training steps = 3905, batch size = 128, max sequence length = 8192 tokens, trainings are performed in bfloat16 precision, learning rate is initially set to $1 \times 10^{-5}$, linearly warms up for 1\% of the total training steps, and decays to $7 \times 10^{-9}$ following a cosine schedule, optimization is performed using the AdamW algorithm with parameters $\beta_1 = 0.9$ and $\beta_2 = 0.95$.

\section{Benchmarks}
\label{ref:Benchmarks}
We conduct our evaluation across two distinct benchmark categories: real-world-level benchmarks and algorithm-level benchmarks. For real-world-level benchmarks we include:

\textbf{BigCodeBench-Instruct} \citep{zhuo2024bigcodebench} is designed to test LLMs’ understanding of human intent in coding. It comprises 1,140 natural language-driven Python tasks—148 of which are categorized as hard. Covering 7 domains and 139 libraries, BCB-Instruct challenges LLMs with real-world programming tasks in data analysis, web development, and beyond. The benchmark emphasizes accurate task understanding, correct decomposition of reasoning steps, and practical problem-solving, making it highly aligned with evaluating LLMs’ real-world code generation abilities.

\textbf{NaturalCodeBench} \citep{zhang-etal-2024-naturalcodebench} is a benchmark designed to reflect the complexity and diversity of real-world programming scenarios, featuring both Python and Java tasks spanning six domains: Front-End Development, Algorithms, Data Science, Artificial Intelligence, Software Engineering, and System Administration. We utilize its open-source development set, which includes 70 Python and 70 Java real-world tasks.

We include the following algorithm-level benchmarks:

\textbf{HumanEval} \citep{chen2021codex} \citep{evalplus} and \textbf{MBPP} \citep{austin2021program} (collectively known as EvalPlus) are benchmarks designed to assess LLMs' capabilities on functional programming tasks. Specifically: HumanEval v0.1.10 consists of 164 hand-written problems, each with a function signature and detailed docstrings, to evaluate LLMs’ proficiency in Python code completion. MBPP v0.2.0 contains 378 problems with natural language instructions, focusing on assessing LLMs' code generation abilities.

\textbf{LiveCodeBench} \citep{jain2024livecodebench} gathers high-quality tasks from competitive programming contests such as LeetCode, AtCoder, and CodeForces to rigorously evaluate code generation skills. For our evaluation, we selected 279 LiveCodeBench tasks created between August 2024 and January 2025 to minimize potential data contamination.

\section{Baseline Details}
\label{ref:Baseline Details}
We conduct a comprehensive comparison with open-source state-of-the-art models in the size of approximately 1.5B and 7B. Specifically, for the 1.5B parameter category, the baseline models include: Deepseek-Coder-1.3B-Instruct \citep{Guo2024DeepSeekCoderWT}, and Qwen2.5-Coder-1.5B-Instruct \citep{hui2024qwen25codertechnicalreport}. In the 7B parameter category, the models considered are: OpenCoder-8B-Instruct \citep{Huang2024OpenCoderTO}, CodeQwen1.5-7B-Chat \citep{qwen}, CodeLlama-7b-Instruct \citep{rozière2024codellamaopenfoundation}, Deepseek-Coder-6.7B-Instruct \citep{Guo2024DeepSeekCoderWT}, and Qwen2.5-Coder-7B-Instruct \citep{hui2024qwen25codertechnicalreport}. In addition, we include several general 7B large language models for comparison purposes: Qwen2.5-7B-Instruct \citep{qwen2.5}, and Qwen3-8B \citep{yang2025qwen3technicalreport}. For inference, we set the sampling temperature to 0, use a maximum output length of 2,048 tokens, and report the pass@1 accuracy across all benchmarks to evaluate model performance.

\end{document}